\documentclass[aps,prl,twocolumn,superscriptaddress]{revtex4-1}
\usepackage{graphicx}
\usepackage{latexsym}
\usepackage{amssymb}
\usepackage{amsmath}
\usepackage{amsfonts}
\usepackage{upgreek}
\usepackage{subfigure}
\usepackage{bm}
\usepackage{bbold}
\usepackage{verbatim}
\usepackage[unicode=true,
 bookmarks=false,
 breaklinks=false,pdfborder={0 0 1},backref=false,colorlinks=true]
 {hyperref}
\hypersetup{
 linkcolor=[rgb]{0,0,1},citecolor=[rgb]{0,0,1},urlcolor=[rgb]{0,0,1}}
\usepackage{multirow}
\usepackage{color}
\usepackage{comment}
\usepackage{xcolor}
\usepackage{soul}
\usepackage{tikz}
\usepackage{tabularx}

\renewcommand{\H}{\hat{\mathcal{H}}}
\renewcommand{\a}{\hat{a}}
\newcommand{\ad}{\hat{a}^\dagger}

\newcommand{\n}{\hat{n}}

\renewcommand{\ij}{\langle i,j \rangle}

\usepackage{braket}

\begin{document}

\title{Antiferromagnetic bosonic $t$\,--\,$J$ models \\ and their quantum simulation in tweezer arrays}

\author{Lukas~Homeier}
\email{lukas.homeier@physik.uni-muenchen.de}
\affiliation{Department of Physics and Arnold Sommerfeld Center for Theoretical Physics (ASC), Ludwig-Maximilians-Universit\"at M\"unchen, Theresienstr. 37, M\"unchen D-80333, Germany}
\affiliation{Munich Center for Quantum Science and Technology (MCQST), Schellingstr. 4, M\"unchen D-80799, Germany}
\affiliation{ITAMP, Harvard-Smithsonian Center for Astrophysics, Cambridge, Massachusetts 02138, USA}
\affiliation{Department of Physics, Harvard University, Cambridge, Massachusetts 02138, USA}

\author{Timothy~J.~Harris}
\affiliation{Department of Physics and Arnold Sommerfeld Center for Theoretical Physics (ASC), Ludwig-Maximilians-Universit\"at M\"unchen, Theresienstr. 37, M\"unchen D-80333, Germany}
\affiliation{Munich Center for Quantum Science and Technology (MCQST), Schellingstr. 4, M\"unchen D-80799, Germany}

\author{Tizian~Blatz}
\affiliation{Department of Physics and Arnold Sommerfeld Center for Theoretical Physics (ASC), Ludwig-Maximilians-Universit\"at M\"unchen, Theresienstr. 37, M\"unchen D-80333, Germany}
\affiliation{Munich Center for Quantum Science and Technology (MCQST), Schellingstr. 4, M\"unchen D-80799, Germany}

\author{Sebastian~Geier}
\affiliation{Physikalisches Institut, Universit\"at Heidelberg, Im Neuenheimer Feld 226, 69120 Heidelberg, Germany}\affiliation{Department of Physics, Harvard University, Cambridge, Massachusetts 02138, USA}

\author{Simon~Hollerith}
\affiliation{Department of Physics, Harvard University, Cambridge, Massachusetts 02138, USA}

\author{Ulrich~Schollw\"ock}
\affiliation{Department of Physics and Arnold Sommerfeld Center for Theoretical Physics (ASC), Ludwig-Maximilians-Universit\"at M\"unchen, Theresienstr. 37, M\"unchen D-80333, Germany}
\affiliation{Munich Center for Quantum Science and Technology (MCQST), Schellingstr. 4, M\"unchen D-80799, Germany}

\author{Fabian~Grusdt}
\affiliation{Department of Physics and Arnold Sommerfeld Center for Theoretical Physics (ASC), Ludwig-Maximilians-Universit\"at M\"unchen, Theresienstr. 37, M\"unchen D-80333, Germany}
\affiliation{Munich Center for Quantum Science and Technology (MCQST), Schellingstr. 4, M\"unchen D-80799, Germany}

\author{Annabelle~Bohrdt}
\email{annabelle.bohrdt@physik.uni-regensburg.de}
\affiliation{Institute of Theoretical Physics, University of Regensburg, Regensburg D-93053, Germany}
\affiliation{Munich Center for Quantum Science and Technology (MCQST), Schellingstr. 4, M\"unchen D-80799, Germany}
\affiliation{ITAMP, Harvard-Smithsonian Center for Astrophysics, Cambridge, Massachusetts 02138, USA}
\affiliation{Department of Physics, Harvard University, Cambridge, Massachusetts 02138, USA}

\date{\today}
\begin{abstract}
The combination of optical tweezer arrays with strong interactions -- via dipole-exchange of molecules and van-der-Waals interactions of Rydberg atoms -- has opened the door for the exploration of a wide variety of quantum spin models. A next significant step will be the combination of such settings with mobile dopants: This will enable to simulate the physics believed to underlie many strongly correlated quantum materials. Here we propose an experimental scheme to realize bosonic $t$\,--\,$J$ models via encoding the local Hilbert space in a set of three internal atomic or molecular states. By engineering antiferromagnetic (AFM) couplings between spins, competition between charge motion and magnetic order similar to that in high-$T_c$ cuprates can be realized. Since the ground states of the $2$D~bosonic AFM~$t$\,--\,$J$~model we propose to realize have not been studied extensively before, we start by analyzing the case of two dopants -- the simplest instance in which their bosonic statistics plays a role, and contrast our results to the fermionic case. We perform large-scale density matrix renormalization group (DMRG) calculations on six-legged cylinders, and find a strong tendency for bosonic holes to form stripes. This demonstrates that bosonic, AFM $t$\,--\,$J$~models may contain similar physics as the collective phases in strongly correlated electrons.
\end{abstract}
\maketitle

\begin{figure}[t!!]
\centering
\includegraphics[width=\linewidth]{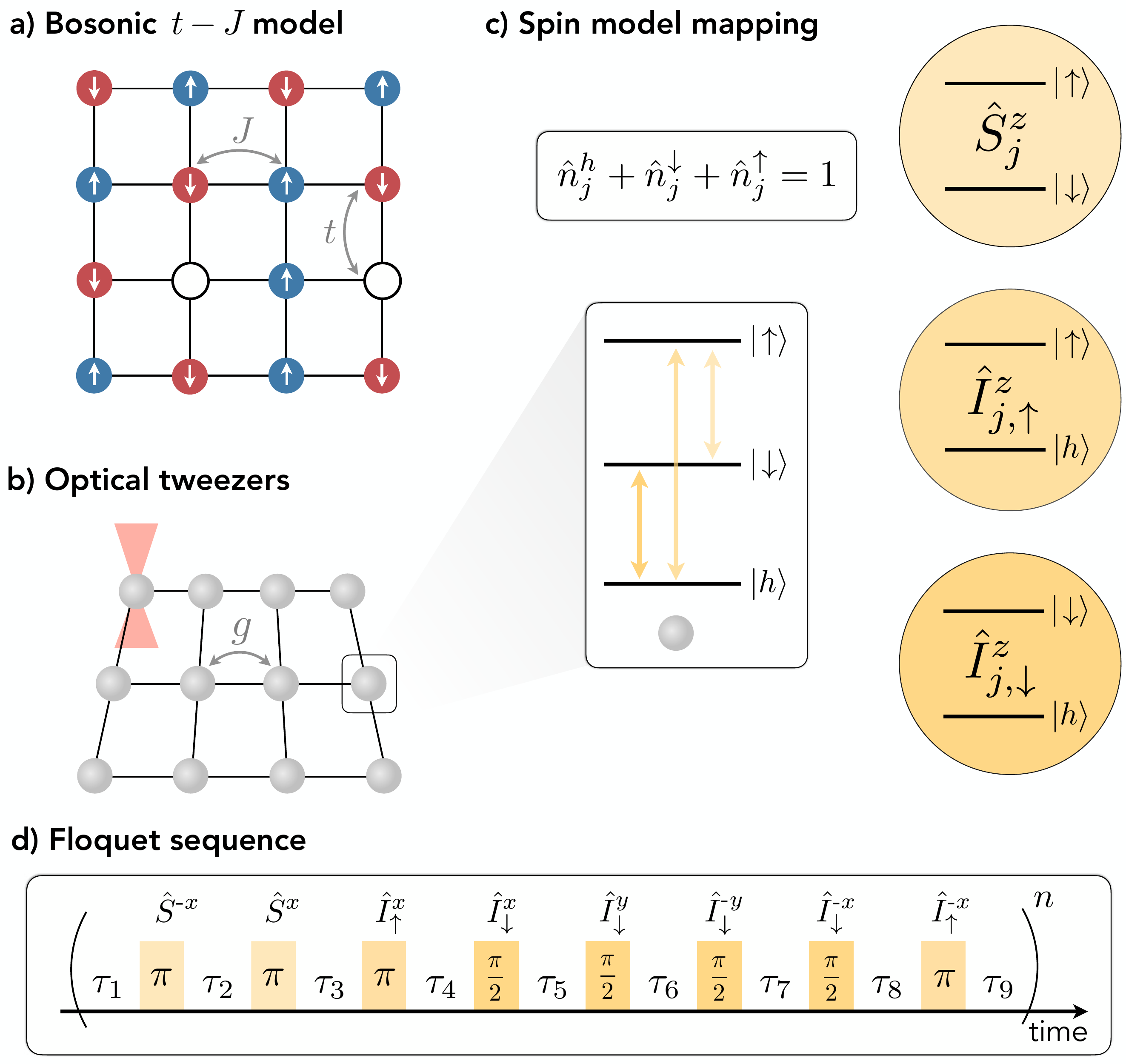}
\caption{\textbf{Schwinger boson mapping.} \textbf{a)}~The $t$\,--\,$J$~model describes hopping of spin-1/2 particles on a lattice with tunneling amplitude~$t$  together with magnetic interactions~$J$. For AFM interactions, the motion of holes~($\ket{h}$) or particles~$(\ket{\downarrow}, \ket{\uparrow})$ frustrates the spin order yielding rich physics for both fermionic and bosonic particles. \textbf{b)}~The latter can be implemented in the internal states of ultracold molecules or Rydberg atoms spatially localized in optical lattices or on an arbitrary graph of tweezer arrays. \textbf{c)}~The local Hilbert space~$\{ \ket{h}, \ket{\downarrow}, \ket{\uparrow} \}$ can be encoded in the internal rotational states of molecules and we define the three Schwinger spins~$\hat{\mathbf{S}},\,\hat{\mathbf{I}}_{\uparrow}$ and~$\hat{\mathbf{I}}_{\downarrow}$, which allows us to exactly represent the $t$\,--\,$J$~Hamiltonian as a spin model. \textbf{d)}~In the isolated three level subspace of rotational states~$\ket{N}$ with $N=0,1,2$, the molecular Hamiltonian we consider has XY interactions between~$N=0,1$. By performing periodic rotations on the~$\hat{\mathbf{S}}$- and~$\hat{\mathbf{I}}_\sigma$-Bloch spheres, the effective Floquet Hamiltonian in spin representation can be engineered. The duration~$\tau_n$ of individual Floquet evolution steps determines the effective coupling strengths of the target Hamiltonian~\eqref{eq:tJ-XXZ} (here with~$V=0$). }
\label{fig1}
\end{figure}

%%%%%%%%%%%%%%%%%%%%%%%%%%%%%%%%%%
\textbf{Introduction.---}Trapping, manipulating and controlling individual qubits in optical tweezer arrays~\cite{Endres2016, Anderegg2019,Kaufman2021,Browaeys2020,Holland2023,Zhang2022} has enabled the observation of intriguing many-body physics. 
Rydberg tweezer platforms stand out with their large Ising or dipolar interaction strengths~\cite{Leseleuc2019,Chen2023}, while cold molecules remain coherent for seconds~\cite{Gregory2021} and offer an entire ladder of rotational states~\cite{Sundar2018}. 
So far, experiments have demonstrated a variety of equilibrium~\cite{Semeghini2021,Ebadi2021} and dynamical phenomena~\cite{Yan2013,Bernien2017,Keesling2019,Christakis2023} of quantum magnetism. 
For example, the toolkit of strong interactions, geometric frustration and novel readout of non-local correlators have revealed topological spin liquid order in Rydberg tweezer arrays~\cite{Verresen2021,Semeghini2021}.

One goal of analog quantum simulators is to develop our understanding of the microscopic mechanisms underlying strong correlated quantum matter.
Combining spin models with physical tunneling~$t$ of particles~\cite{Kaufman2014} yields \textit{doped quantum magnets}, where mobile dopants frustrate magnetic order~\cite{Auerbach1994} and the statistics of the particles plays a crucial role.
Due to its intimate connection to strongly correlated electrons, much effort has been invested in the exploration and quantum gas microscopy of the Fermi-Hubbard model~\cite{Keimer2015, Bohrdt2021} with on-site interaction~$U$, using ultracold atoms in optical lattices~\cite{Cheuk2016,Mitra2017,Mazurenko2017,Koepsell2019,Xu2023a}.
The underlying superexchange mechanism naturally leads to AFM interactions~$J = 4t^2/U$ in fermionic systems, while bosonic models have effective ferromagnetic interactions~\cite{Auerbach1994,Duan2003}.

The behaviour of bosonic holes doped into an AFM background raises several interesting questions, but has so far remained elusive due to the ferromagnetic interactions in spin-$1/2$ Bose-Hubbard models.
For example, the microscopic mechanism of hole pairing might not be specific to the Fermi-Hubbard model but instead a universal feature of a broad class of related systems with strong spin-charge correlations; such as the model discussed below.

In this Letter, (i) we study a model combining AFM spin models with mobile (hardcore) bosonic hole dopants in one or two spatial dimensions and (ii) we propose experimental schemes realizing this scenario, suitable for implementation in systems of ultracold polar molecules or Rydberg atoms, where the hole dopants are encoded in the internal degrees-of-freedom.

In particular, we map a bosonic $t$\,--\,$J$~model~\cite{Smakov2004,Boninsegni2008,Aoki2009,Nakano2011,Nakano2012,Sun2021,Jepsen2021} onto a pure spin model comprised of three Schwinger bosons, which can be implemented hardware-efficiently using the Floquet technique in tweezer systems with dipolar or van-der-Waals interactions.
Here, the system time evolves under its natural e.g.\,XY~interactions followed by a specific sequence of rotations within the three internal states.
The effective dynamics of the bosonic excitations is then governed by a $t$\,--\,$J$~Hamiltonian.
The tunability enables us to engineer regimes that have not been accessible before, e.g. AFM $\mathrm{XXZ}$~interactions, $J>t$, explicit hole-hole (anti)binding potentials or randomized interactions~\cite{Christos2022}.

With the deterministic loading and preparation of product states, time evolution under a tunable Hamiltonian in arbitrary geometry, and ultimately readout by snapshots in the Fock basis, we present a realistic experimental protocol to probe doped bosonic quantum magnets.
We highlight the relevance of $2$D~\textit{bosonic AFM} $t$\,--\,$J$~models by calculating the ground state of a six-legged cylinder doped with two holes using DMRG.
We then compare these results with those obtained from the traditional fermionic $t$\,--\,$J$~model.

%%%%%%%%%%%%%%%%%%%%%%%%%%%%%%%%%%
\textbf{Bosonic $t$\,--\,$J$~model as a spin system.---}The main ingredient to realize doped spin models, such as the $t$\,--\,$J$~model shown in Fig.~\ref{fig1}a, is a mapping from the original model onto a new model described by Schwinger bosons.
The new spin model is then suitable for implementions in established experimental platforms and the desired interactions can be engineered using the Floquet driving technique.

The $t$\,--\,$J$~model describes (hardcore) mobile \mbox{spin-$1/2$} particles on a $d$-dimensional lattice with magnetic interactions; hence the local Hilbert space is spanned by the hole and \textit{one} particle states $\{\ket{h}$,\,$\ket{\downarrow}$,\,$\ket{\uparrow}\}$.
Here, we investigate bosonic particles~$\ket{\sigma} = \ad_{j,\sigma}\ket{vac}$, where we express spins in the Schwinger representation~\mbox{$\hat{\mathbf{S}}_j=\frac{1}{2} \sum_{\sigma,\sigma'} \ad_{j,\sigma} \boldsymbol{\tau}_{\sigma, \sigma'} \a_{j,\sigma'}$} with Pauli matrices~\mbox{$\boldsymbol{\tau} = (\tau^x,\tau^y,\tau^z)$} and $\sigma =\,\downarrow, \uparrow$.
Further, we introduce the (hardcore) bosonic hole operator~$\ket{h} = \ad_{j,h}\ket{vac}$.

To obtain the correct Hilbert space, the Schwinger bosons have to fulfill the local constraint
\begin{align} \label{eq:number_constraint}
     \n_j^h + \n_j^{\downarrow} + \n_j^{\uparrow} = 1,
\end{align}
where $\n_j^\sigma = \ad_{j,\sigma}\a_{j,\sigma}$ and $\n_j^h = \ad_{j,h}\a_{j,h}$ are the local spin and hole densities, respectively, see Fig.~\ref{fig1}c.

The bosonic $t$\,--\,$J$~Hamiltonian is given by
\begin{align}
\begin{split} \label{eq:tJ-general}
    \H_{t-J} &= -\sum_{i<j} t_{ij} \sum_{\sigma} \left( \a^\dagger_{i,\sigma}\a_{i,h}\a^\dagger_{j,h}\a_{j,\sigma} + \mathrm{h.c.}  \right) \\
    &+ \sum_{i<j}\sum_{\alpha} J^\alpha_{ij}\, \hat{S}^\alpha_i\hat{S}^\alpha_j  + \sum_{i<j}  V_{ij}  \n^h_i \n^h_j,
\end{split}
\end{align}
with~$\alpha=x,y,z$ and the couplings can have arbitrary connectivity and range.
The first term~$\propto t$ describes tunneling of particles, the second term~$\propto J^{\alpha}$ describes magnetic XXZ interactions with~$J^{x}=J^{y} = J^{\perp}$, and the last term~$\propto V$ is a hole-hole interaction.

The model~(\ref{eq:tJ-general}) gains its importance because it captures the low-energy effective theory of the repulsive Fermi- or Bose-Hubbard models in the strong coupling regime~\mbox{$U \gg t$}~\cite{Auerbach1994}.
However, the perturbative derivation exactly determines the couplings, which for nearest-neighbour (NN) hopping are given by~$J^\alpha = \pm 4t^2/U$ and~$V = -J(\pm 2 - 1)/4$ for the fermionic~$(+)$ or bosonic~$(-)$ models, respectively.

Our proposed scheme for realizing the model~\eqref{eq:tJ-general} in experiment enables broad tunability~\cite{Gorshkov2011,Coulthard2017} of the Hamiltonian parameters.
In particular, the ability to tune the ratio between the hole-hole interaction~$V$ and magnetic interactions~$J$ in our model facilitates exploration of potentially interesting pairing regimes, which we study numerically in the second part of this paper. 

First, we perform an exact mapping of Hamiltonian~\eqref{eq:tJ-general} onto a new XXZ spin model comprised of the three spin-$1/2$ Schwinger spins~$\hat{\mathbf{S}}_j$, $\hat{\mathbf{I}}_{j,\downarrow}$ and $\hat{\mathbf{I}}_{j,\uparrow}$ with
\begin{align}
\begin{split} \label{eq:SchwingerBosons}
        \hat{S}^{z}_{j} &= \frac{1}{2} \left(  \n^{\uparrow}_j - \n^{\downarrow}_j \right) ~~~~~ \hat{S}^{+}_{j} = \ad_{j,\uparrow}\a_{j,\downarrow}\\
        \hat{I}^{z}_{j,\sigma} &= \frac{1}{2} \left(  \n^{\sigma}_j - \n^{h}_j \right)  ~~~~~ \hat{I}^{+}_{j,\sigma} = \ad_{j,\sigma}\a_{j,h},
\end{split}
\end{align}
from which we obtain (up to a constant energy shift)
\begin{align}
\begin{split} \label{eq:tJ-XXZ}
    \H_{t-J} &= -\sum_{i<j} \sum_{\alpha, \sigma} t^\alpha_{ij} \hat{I}^{\alpha}_{i,\sigma}\hat{I}^{\alpha}_{j,\sigma} 
    + \sum_{i<j}\sum_{\alpha} g^\alpha_{ij}\, \hat{S}^\alpha_i\hat{S}^\alpha_j.
\end{split}
\end{align}
We neglect a chemical potential term for the holes since we assume the total number of particles is conserved. 
The form of Eq.~\eqref{eq:tJ-XXZ} is very useful for our proposed implementation below, but we emphasize that the Schwinger spins are not mutually independent, i.e.~$[ \hat{S}^z_j, \hat{I}^{\pm}_{j,\sigma}] \neq 0$.

The hole-hole interaction renormalizes the XXZ models and we find the following couplings related to Eq.~\eqref{eq:tJ-general} and~\eqref{eq:tJ-XXZ}:
\begin{align}
\begin{split}
    &t_{ij}^x = t_{ij}^y =\dfrac{1}{2}t_{ij} ~~~~~~~~~~~~~~~ t_{ij}^z = -\dfrac{8}{9}V_{ij} \\
    &g_{ij}^{x}=g_{ij}^{y} = J_{ij}^{\perp} ~~~~~~~~~~~~ g_{ij}^z=J_{ij}^z-\dfrac{4}{9}V_{ij}.
\end{split}
\end{align}
So far, we have performed exact transformations and re-written the $t$\,--\,$J$~model in terms of Schwinger bosons.
The Schwinger spins have to fulfill the the number constraint~\eqref{eq:number_constraint}, which induces highly non-trivial spin-charge correlations and thus is beyond a simple spin-1/2 chain.
Likewise, the construction can be formulated in terms of mutually hard-core bosonic statistics, i.e.\, $\a^\dagger_{j,\Sigma}\a^\dagger_{j,\Sigma'}=0$ for~$\Sigma=\downarrow,\uparrow,h$.

The constraint can be elegantly implemented in a spin-1 manifold in e.g.\,ultracold molecule or Rydberg tweezer arrays.
To this end, we propose two schemes, which either utilize dipolar spin exchange interactions to engineer the desired dynamics by Floquet driving, or directly enables the realization of Hamiltonian~\eqref{eq:tJ-XXZ} in three isolated Rydberg states.

%%%%%%%%%%%%%%%%%%%%%%%%%%%%%%%%%%
\textbf{Experimental proposal: Ultracold molecules.---}Ultracold polar molecules have recently demonstrated the realization of an anisotropic XXZ~model in a qubit subspace of rotational states~\cite{Christakis2023}, which can be achieved by Floquet engineering.
There, the system consecutively time evolves under the resonant dipole-dipole interactions followed by fast qubit rotations, i.e. driving microwave transitions between rotational states.

Here, we extend the scheme by using three states in the rotational manifold~$\ket{N}$ with~$N=0,1,2$ and we identify the molecular states $\{ \ket{0}, \ket{1}, \ket{2} \}$ with the local Hilbert space $\{ \ket{h}, \ket{\downarrow}, \ket{\uparrow} \}$ of the $t$\,--\,$J$~model.
The molecular Hamiltonian expressed in terms of the Schwinger spins~\eqref{eq:SchwingerBosons} is given by
\begin{align} \label{eq:molHam}
    \H_\mathrm{mol} = \sum_{i<j} \chi_{ij} \left(  \hat{I}^{+}_{i,\downarrow} \hat{I}^{-}_{j,\downarrow} + \mathrm{h.c.} \right),
\end{align} 
with $\chi_{ij} = \chi (1-3\cos^2{\theta_{ij}})/|\mathbf{r}_{ij}|^3$.
Here, $\mathbf{r}_{ij}$ is the vector connecting lattice sites~$i$ and~$j$, and $\theta_{ij}$ is the angle between the quantization axis and $\mathbf{r}_{ij}$.
In the following, we choose~$\theta_{ij}=\pi/2$ and set the NN distance to~$r_{ij} \equiv 1$.
The XY~coupling strength~$\chi$, is determined by the resonant dipole moments of the molecule~\cite{Wall2015,Sundar2018}, and here we only consider interactions between~$N=0,1$~\cite{Park2023}, while the state~$N=2$ is non-interacting. This can be achieved by using the selection rules~$\Delta m_N= 0, \pm 1$ of the dipole interactions, e.g. we propose to use~$\ket{N=0,m_N=0}$, $\ket{N=1,m_N=0}$ and~$\ket{N=2,m_N=-2}$.

Next, we describe a scheme to realize a $t$\,--\,$J$~model with tunable XXZ~magnetic interactions.
To this end, we consider the molecular Hamiltonian~\eqref{eq:molHam} with flip-flop interactions~$\chi_{ij}$.
By comparing this model to the $t$\,--\,$J$~Hamiltonian~\eqref{eq:tJ-XXZ}, we find that the microscopic model corresponds to a $t$\,--\,$J$~model with tunneling of $\ket{\downarrow}$-particles only.
Hence, we propose to perform consecutive, fast rotations between all pairs of states, i.e. on the $\mathbf{\hat{I}}_\sigma$- and~$\mathbf{\hat{S}}$-Bloch spheres, to obtain a time-averaged Hamiltonian with equal strength $\ket{\uparrow}$- and $\ket{\downarrow}$-particle tunneling.
The sequence of Floquet rotations is shown in Figure~\ref{fig1}d:
Tuning the times~$\tau_n$ of Floquet steps allows for the implementation of models with tunable ratios~$-t/J^z>0$ and~$J^\perp/J^z>0$ ($J_z>0$) and~$V=0$.
This result holds in first-order Floquet theory and is derived in the Supplementary Material; a comparison between the target $t$\,--\,$J$~model and Floquet time evolution using exact diagonalization shows excellent agreement and demonstrates the robustness of our proposed scheme.

We emphasize that the long-range interactions directly transfer to the effective model and hence a $t$\,--\,$J$~model with $r^{-3}$ tails is realized in cold molecules.
Enriching the Floquet protocol by spatial rearrangements~\cite{Bluvstein2022}, pure NN interactions or even models with arbitrary connectivity can be implemented in principle.
Depending on the stability of DC electric fields, the fidelity of microwave transitions and coherence times across multiple rotational levels~\cite{Ni2018,Rosenband2018}, effective Floquet Hamiltonians~\cite{Sun2023} of differing complexity can be realized; in particular in the Supplementary Material we present a Floquet sequence that gives rise to $t$\,--\,$J$\,--\,$V$~models with~$0 < V/J^z < 9/4$ and~$J^\perp/J^z > 1/2$.

%%%%%%%%%%%%%%%%%%%%%%%%%%%%%%%%%%
\textbf{Experimental proposal: Rydberg tweezer arrays.---}Rydberg atoms in optical lattices~\cite{Zeiher2016} and tweezer arrays have become an established platform in the quantum simulation of magnetism ~\cite{Leseleuc2019,Chen2023,Semeghini2021,Ebadi2021,Bernien2017,Keesling2019}.
In particular, tunable spin-$1/2$ XXZ models have previously been realized via Rydberg dressing~\cite{Steinert2023}, Floquet engineering~\cite{Geier2021,Scholl2022} and precise selection of Rydberg states~\cite{Franz2022_arxiv}.

The proposed model~\eqref{eq:tJ-XXZ} requires control over interactions within a three-level system.
We propose a direct implementation within three Rydberg states by identifying
\begin{align}
\ket{nS} = \ket{\downarrow} ~~~~ \ket{n'P} = \ket{h} ~~~~ \ket{n''S} = \ket{\uparrow}.
\end{align}
The resonant dipolar exchange between states of different parity implements long-ranged tunnelings, e.g., the exchange interaction between the pair of atoms $\ket{nS_i,n'P_j} \leftrightarrow \ket{n'P_i,nS_j}$ at site~$i$ and~$j$ corresponds to tunneling of $\downarrow$-particles with amplitude~$t^\downarrow_{ij} \propto r_{ij}^{-3}$. Models with either approximate ${\rm SU}(2)$~invariant tunnelings or spin-dependent tunnelings can be implemented.

Further, the spin interactions can be induced by choosing suitable pair states~$\ket{nS,n''S}$, such that the resulting van-der-Waals interactions~$\propto r^{-6}$ give rise to flip-flop~$J_\perp$ and Ising terms~$J_z$~\cite{Whitlock2017}, as demonstrated in Ref.~\cite{Franz2022_arxiv}.
Notably, the different scaling behaviours of the tunneling and spin interactions allows us to tune the ratio of~$t/J$ over a wide range. 

In the (exact) mapping from the Rydberg Hamiltonian onto the spin model, additional terms appear from the anisotropy in the diagonal van-der-Waals interaction between pair states. These terms give rise to a $t$\,--\,$J$\,--\,$V$\,--\,$W$~model, see Eq.~\eqref{eq:tJ-general}, with spin-hole interactions~$\propto W_{ij} \hat{n}^h_{i}\hat{S}^z_j $. Using the states from Ref.~\cite{Franz2022_arxiv}, we find the latter to be negligible.

\begin{figure}[t!!]
\centering
\includegraphics[width=\linewidth]{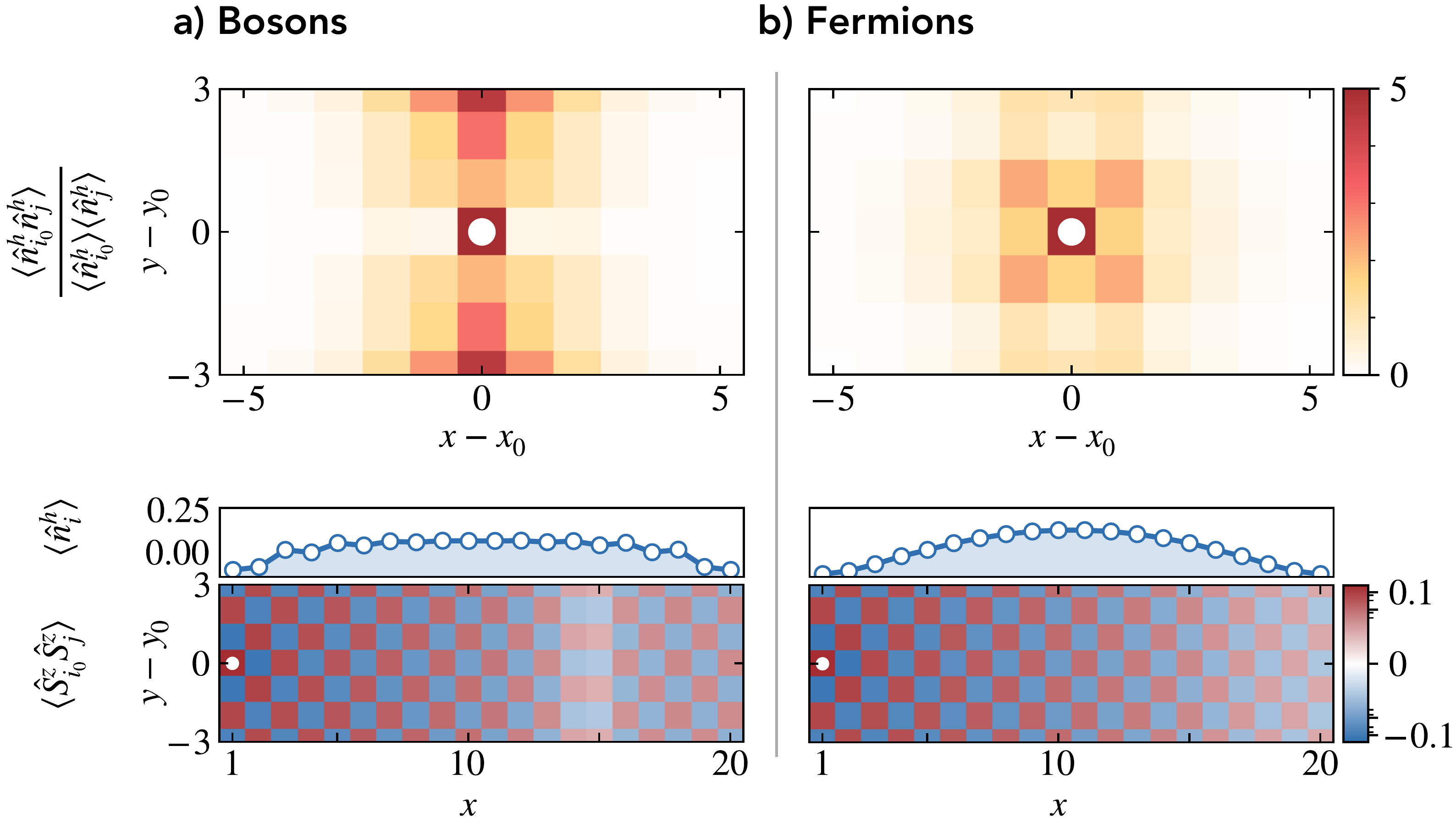}
\caption{\textbf{Bosonic vs. fermionic two-hole states.} We show ground-state correlation functions for a~$t$\,--\,$J$~model with two \textbf{a)} bosonic and \textbf{b)} fermionic holes obtained from DMRG calculations on $20\times 6$~cylinders. We find distinctly different behaviour for bosons and fermions by evaluating the hole-hole (top) and spin-spin correlation functions (bottom). \textbf{a)} The hole-hole correlator, centered around its reference site, $\bm{i}_0 = (x_0=10, y_0=3)$, shows a charge-density wave pattern around the short direction of the cylinder. Additionally, we find a domain wall in the spin-spin correlation function with reference site $\bm{i}_0 = (x_0=1, y_0=3)$, on the boundary (white dot); hence the bosons form a stripe.  \textbf{b)} The fermions are tightly bound into an isotropic pair embedded in a homogeneous AFM background. For both statistics, we plot the hole density averaged around the short direction of the cylinder, which also serves as a marker for convergence of our results (see Supplementary Material). }
\label{fig2}
\end{figure}

%%%%%%%%%%%%%%%%%%%%%%%%%%%%%%%%%%
\textbf{Spin-charge order in the bosonic $t$\,--\,$J$ model.---}Understanding the nature of mobile dopants in strongly correlated phases of matter has a long history, motivated by high-$T_c$ superconductors and more recently by layered 2D materials~\cite{Cao2018}.
The fate of the AFM Mott insulator under doping is still debated; however, experiments in cuprates have revealed that even a few percent of fermionic dopants can lead to a robust d-wave superconducting ground state~\cite{Keimer2015,Xu2023}.
Hence, strong pairing of charge carriers -- the hole dopants -- mediated by magnetic interactions~\cite{OMahony2022} likely plays a key role.

Here, we perform a first numerical study of hole dopants in the ground state of the $2$D~bosonic AFM $t$\,--\,$J$~model, comparing our results to an equivalent calculation using the standard fermionic $t$\,--\,$J$~model.
Let us emphasize that previous studies of the bosonic $t$\,--\,$J$~model have considered either lower dimensions, high temperature expansions or \textit{partial} AFM couplings ($J^z>0, J^\perp \leq 0$)~\cite{Smakov2004,Boninsegni2008,Aoki2009,Nakano2011,Nakano2012,Sun2021,Jepsen2021}.
In contrast, our model takes a further step towards strongly correlated materials by studying fully antiferromagnetic interactions in the spin sector~($J^z, J^\perp > 0$) with the cost of introducing a sign problem at low temperatures; there our model is intractable for large-scale quantum Monte Carlo simulations.

To this end, the ground state with two holes in the zero-magnetization sector, $\hat{S}_z=0$, of the SU$(2)$-invariant version of Eq.~\eqref{eq:tJ-general},~$J^\alpha \equiv J$, was obtained from DMRG calculations on a long cylinder \cite{White1992,Schollwoeck2011,Hubig2017,Syten}; the interactions are restricted to NN and have strength~$t/J=2$ and~$V/J=-1/4$ (see Supplementary Material).
To analyze the structure of the obtained pair wavefunctions, we extract (i) the reduced hole-hole correlation~$\langle \hat{n}^h_i \hat{n}^h_j \rangle$/$\langle \hat{n}^h_i \rangle \langle \hat{n}^h_j \rangle$ and (ii) the spin-spin correlation~$\langle \hat{S}^z_i \hat{S}^z_j \rangle$ functions shown in Fig.~\ref{fig2}.

The well-known case of fermionic holes~\cite{Qin2020,Xu2023,Lu2023} indicates the formation of a tightly bound pair state, which can be seen from the C$4$-symmetric hole-hole correlations (Fig.~\ref{fig2}b, top) and the absence of a spin domain wall across the hole-rich region (Fig.~\ref{fig2}b, bottom).
While the intuitive picture for bosons suggests the holes to condense and similarly bunch together, we find a surprising situation: the bosons have a tendency towards stripe formation.
At finite density of holes, such stripes form in e.g. cuprate materials, describing periodic charge modulations bound to $\pi$-phase shifts of the spin-spin correlations (domain walls) across the hole-rich regions.

In our small system simulation, the two bosonic holes show strong tendency to pair along the short, periodic direction of the cylinder, as evident from the hole-hole correlations (Fig.~\ref{fig2}a, top).
In contrast to the fermionic case, we observe a spin domain wall across the hole-rich region (Fig.~\ref{fig2}a, bottom), a hallmark of stripe formation.
Additionally, the charge correlations show short-range repulsion along the short direction, distinctly different from the structure of the C$4$-invariant pair of the fermionic holes but resembling the situation in a stripe.
Both scenarios, tightly bound pairs and stripe correlations, are marking phases observed in strongly correlated electrons~\cite{Wietek2021,Xu2023}.

This minimal instance -- comparing two-hole fermionic and bosonic states -- already shows rich phenomenology and demonstrates an intriguing first experimental application of our proposal.
Future experimental and numerical studies of the bosonic AFM $t$\,--\,$J$~model can be expected to provide a fresh perspective from which to advance our current understanding of the physics of doped Mott insulators.

%%%%%%%%%%%%%%%%%%%%%%%%%%%%%%%%%%
\textbf{Experimental probes.---}In the following, we discuss observables that can be obtained from snapshots in our proposed experimental scheme.
From low- to high-doping, we suggest a number of useful probes: single-hole angle-resolved photoemission spectroscopy (ARPES), binding energies, $g^{(2)}$ correlation functions and transport.

A single hole dopant in an AFM background forms a quasiparticle, the magnetic polaron, with rich internal structure~\cite{Grusdt2018}.
Previous experiments focused on measurements of hole-spin-spin correlation functions~\cite{Koepsell2019} or dynamical probes~\cite{Ji2021}.
However, a plethora of ro-vibrational excitation modes are predicted from numerical calculations~\cite{Bohrdt2021a}, which could be revealed by single-hole ARPES~\cite{Brown2019}.
In our proposed model, a momentum-insensitive yet spin-selective measurement can be easily performed by driving microwave transitions between the $\ket{\downarrow}$ ($\ket{\uparrow}$) and~$\ket{h}$ states, in order to evaluate the transition probabilities.
We note that in the $t-J^z$~model ($J^\perp=0$) the energy is approximately momentum independent; in our proposed scheme this can be implemented as a limit of the XXZ~model with strong Ising terms.
Moreover, the characteristic excitation spectrum of the single hole is a staircase of string-excited states~\cite{Trugman1988,Kane1989,Grusdt2018,Wrzosek2021} with energies scaling as~$t^{1/3}(J^z)^{2/3}$, which could be directly probed by momentum-independent measurements.

In contrast to optical lattice experiments, platforms utilizing tweezer arrays allow for direct access to energies~$\langle \H \rangle$ by taking snapshots in the different bases of~$\hat{S}^\alpha$ and $\hat{I}^\alpha_\sigma$ and evaluating the terms individually in Hamiltonian~\eqref{eq:tJ-XXZ}.
The experimental protocol first requires an adiabatic state preparation protocol, e.g.\, by preparing a deterministic, low-energy N{\'e}el state followed by tuning suitable parameters.
Repeating the experiment for zero, one and two holes, then allows to measure e.g. their binding energy directly.

On the other hand, indirect signatures of pairing or stripe formation can be obtained from $g^{(2)}$~correlation functions.
Moreover, the tunability of our model may give a new perspective on the nature of the ground state.
In particular, tuning the magnetic or hole-hole interaction~$V$ can (un)favour pairing, which manifests in the hole-hole distance.
Moreover, at finite doping we speculate that the phase diagram -- similar to its fermionic counterpart -- hosts instabilities towards incommensurate charge and spin ordered phases of matter.
Hence, the formation (or absence) of stripes at finite doping with its correlation between density and spin domain walls can be investigated from state-dependent snapshots. 

Lastly, we suggest to probe spin and charge transport by time-evolving an initial product state, e.g.\,a charge-density wave~\cite{Brown2019_Transport}.
The control of atoms on the single-particle level in tweezer arrays~\cite{Chen2023} naturally suggests to apply quench protocols to an initial product state.
Studying transport properties in $2$D~$t$\,--\,$J$~models with long-range interactions and at finite doping could give important insights into exotic phases of matter with potential connections to strange metallicity.

%%%%%%%%%%%%%%%%%%%%%%%%%%%%%%%%%%
\textbf{Discussion \& Outlook.---}We have studied a class of models, where AFM interactions are combined with bosonic hole dopants in 2D.
Our proposed model is of particular interest due to the prospect of near-term realizations in analog quantum simulation experiments.
While we have predominantly discussed schemes for analog quantum simulation, we also envisage future applications in hybrid digital-analog platforms~\cite{GonzalezCuadra2023}, where fast physical tunneling of bosonic or fermionic (Rydberg) atoms can be realized. Further, extensions in other 3-manifolds of circular Rydberg states may be possible~\cite{Kruckenhauser2022}.

We demonstrated the relevance for strongly correlated materials and potential connections to high-$T_c$ superconductivity using state-of-the-art numerical techniques; this motivates future theoretical, numerical and experimental studies of bosonic AFM $t$\,--\,$J$~models.
Additionally, the precise tunability of our proposed model opens exciting new avenues to explore pairing mechanisms and collective phases of doped Mott insulators~\cite{Lee2006}. For example, via tuning the hole-hole interaction to realize a hole binding-unbinding transition.
Lastly, the experimental building block could enable the realization of more elaborate strongly-correlated systems, e.g. non-Abelian lattice gauge theories~\cite{HalimehHomeier2023}.

%%%%%%%%%%%%%%%%%%%%%%%%%%%%%%%%%%
\textbf{Note added.---}During the publication process, we became aware of a tunable, fermionic $t$\,--\,$J$~model implemented in cold molecules using physical tunneling of molecules in a lattice~\cite{Carroll2024}.

\textbf{Acknowledgments.---}We are very grateful to \mbox{K.-K.}~Ni, A.~Park, G.~Patenotte and L.~Picard for fruitful discussions and insights regarding the cold molecule scheme, and we wish to thank \mbox{N.-C.}~Chiu for valuable feedback on Rydberg atom implementations. We thank D.~Wei and J.~Zeiher for insightful discussions. L.H. acknowledges support from the Studienstiftung des deutschen Volkes. T.J.H. acknowledges funding by the Munich Quantum Valley (MQV) doctoral fellowship program, which is supported by the Bavarian state government with funds from the Hightech Agenda Bayern Plus. S.H. acknowledges funding through the Harvard Quantum Initiative Postdoctoral Fellowship in Quantum Science and Engineering. This research was funded by the Deutsche Forschungsgemeinschaft (DFG, German Research Foundation) under Germany’s Excellence Strategy—EXC-2111—390814868, by the European Research Council (ERC) under the European Union’s Horizon 2020 research and innovation programme (grant agreement number 948141), by the NSF through a grant for the Institute for Theoretical Atomic, Molecular, and Optical Physics at Harvard University and the Smithsonian Astrophysical Observatory.

%% references
%\section*{References}
\bibliographystyle{apsrev4-1}

%merlin.mbs apsrev4-1.bst 2010-07-25 4.21a (PWD, AO, DPC) hacked
%Control: key (0)
%Control: author (72) initials jnrlst
%Control: editor formatted (1) identically to author
%Control: production of article title (-1) disabled
%Control: page (0) single
%Control: year (1) truncated
%Control: production of eprint (0) enabled
%

%% supplement
\pagebreak
\clearpage
\newpage
\appendix
%\onecolumngrid
\widetext
\begin{center}
\textbf{\large Supplemental Materials: Antiferromagnetic bosonic $t$\,--\,$J$ models and their quantum simulation in tweezer arrays}
\end{center}

%%%%%%%%%% Merge with supplemental materials %%%%%%%%%%
%%%%%%%%%% Prefix a "S" to all equations, figures, tables and reset the counter %%%%%%%%%%
\setcounter{equation}{0}
\setcounter{figure}{0}
\setcounter{table}{0}
\setcounter{page}{1}
\makeatletter
\renewcommand{\theequation}{S\arabic{equation}}
\renewcommand{\thefigure}{S\arabic{figure}}

\section{Floquet engineering in ultracold molecules}\label{subsec:Floquet_wIsing}
As discussed in the main text, the $t$\,--\,$J$~model can be rewritten using the Schwinger boson representation,~Eq.~\eqref{eq:tJ-XXZ}, yielding a spin model encoded in a three level system. 
Experimental setups with ultracold molecules have the capability to coherently address $N$-level systems, which interact via long-ranged flip-flop spin interactions between rotational states with~$\Delta N = \pm 1$ and~$\Delta m_N =0 , \pm 1$.
Further, because the dipole moment enables short Rabi pulse times between different $N$-levels with frequencies in the microwave regime, they are highly suitable for Floquet engineering.

Here, we start from the molecular Hamiltonian~$\H_{\mathrm{mol}}$, Eq.~\eqref{eq:molHam}, and derive the effective Floquet Hamiltonian to first order, i.e.\,we neglect terms~$\mathcal{O}(\tau^2/T_F^2)$, where~$\tau$ is duration of a single step within the Floquet cycle of length~$T_F$.
We simulate the dynamics of a building block with three sites to confirm excellent agreement between the exact and the Floquet averaged dynamics.

\begin{figure}[t]
\centering
\includegraphics[width=0.95\linewidth]{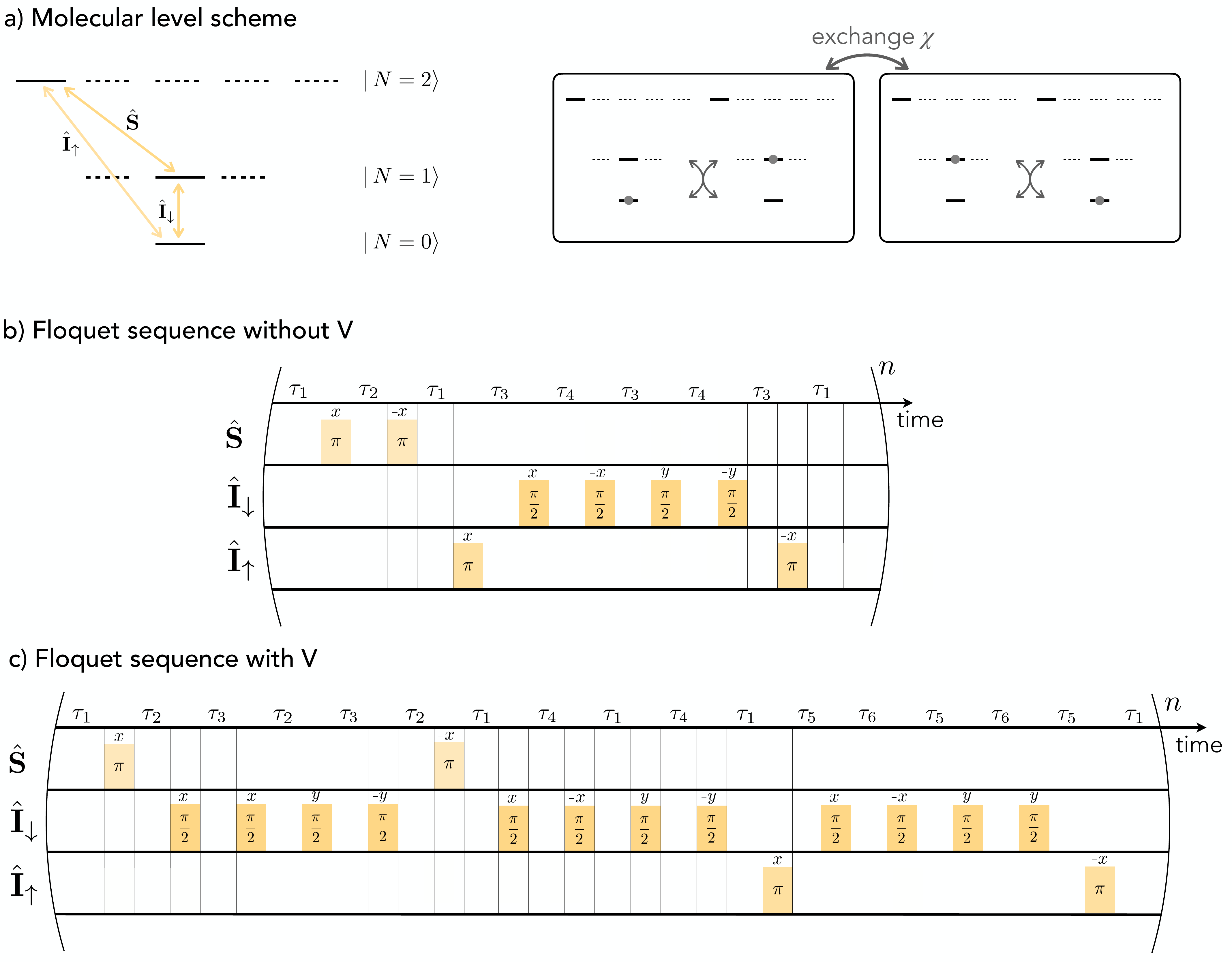}
\caption{\textbf{Molecular scheme and Floquet sequence.} \textbf{a)} We propose to Floquet engineer a $t$\,--\,$J$\,--\,$V$~model starting from a molecular Hamiltonian with two interacting levels (e.g. $\ket{N=0, m_N=0} \leftrightarrow \ket{N=1, m_N=0}$) and an auxiliary non-interacting level (e.g. $\ket{N=2, m_N=-1})$). Using one- and two-photon microwaves with Rabi frequency~$\Omega$ one can rotate between all pairs of levels. To obtain a target $t$\,--\,$J$\,--\,$V$~model, we propose to continuously time evolve the three level system under the application of periodic Rabi pulses. \textbf{b)} The sequence of microwave pulses gives rise to a $t$\,--\,$J$~model ($V=0$) with times~$\tau_n$ determined by Eqs.~\eqref{eq:taus_withoutV}a-e. The time evolution under the shown sequence is demonstrated in Fig.~\ref{figSupp_Floquet}. \textbf{c)} Similarly, we propose a Floquet sequence to realize a $t$\,--\,$J$\,--\,$V$~model with times~$\tau_n$ determined by Eqs.~\eqref{eq:taus_withV}a-f. }
\label{figSupp_Floquet_2level}
\end{figure}

In our molecular Hamiltonian, we assume that only two levels interact, e.g.\,by choosing appropriate sublevels with corresponding selection rules, see Fig.~\ref{figSupp_Floquet_2level}.
This has the advantage that the Floquet sequence enables the realization of a highly tunable $t$\,--\,$J$\,--\,$V$~model.
In particular, we show how to realize target models with any
\begin{align}
\begin{split} \label{eq:tunability1}
    \epsilon^t := \frac{-t}{J^z} > 0,
\end{split}
\end{align}
and
\begin{align}
\begin{split} \label{eq:tunability2}
    &\text{(I)}~~~\epsilon^\perp := \frac{J^\perp}{J^z} > 0 ~~\text{and}~~ \epsilon^V = 0\\
     &\text{(II)}~~~\epsilon^\perp := \frac{J^\perp}{J^z} > 1/2 ~~\text{and}~~ 0 < \epsilon^V := \frac{V}{J^z} <\frac{9}{4}.
\end{split}
\end{align}
For the case of antiferromagentic spin interactions, we choose~$J_z > 0$, which results in models with an overall positive sign for the kinetic energy term~$\propto t$.
Below, we propose two different Floquet sequences to realize models with~$V=0$ and~$V \neq 0$, respectively.
Note that more elaborate Floquet sequences, including interactions between all three levels, can be derived but lead to restricted tunability. Moreover, we can achieve parameter regimes with negative couplings by combining the Floquet scheme with spatial rearrangement and anisotropic dipolar interactions.

Our Floquet sequence requires rotations between all three Schwinger spins~$\hat{\mathbf{S}}$, $\hat{\mathbf{I}}_\downarrow$ and $\hat{\mathbf{I}}_\uparrow$ [Eq.~\eqref{eq:SchwingerBosons}].
The non-interacting levels have, by construction, a vanishing transition dipole matrix element but two-photon transitions can be efficiently implemented; therefore one and two-photon microwave transitions can fully rotate within the three level system.

To derive the Floquet sequence, we find it convenient to re-write the target~$t$\,--\,$J$\,--\,$V$~Hamiltonian as
\begin{align}
\begin{split} \label{eq:tJ-withXi}
    \H_{t-J} = \sum_{i<j} \Bigg\{ -&\sum_{\sigma} 2 t_{ij} \left[ \hat{I}^{x}_{i,\sigma}\hat{I}^{x}_{j,\sigma} + \hat{I}^{y}_{i,\sigma}\hat{I}^{y}_{j,\sigma} \right] + J^\perp_{ij} \left[ \hat{S}^x_i\hat{S}^x_j + \hat{S}^y_i\hat{S}^y_j \right] \\
    + &\sum_{\sigma}\frac{4}{9} V_{ij}\left[ (2-\xi) \hat{I}^{z}_{i,\sigma}\hat{I}^{z}_{j,\sigma} + \xi \hat{I}^{z}_{i,\sigma}\hat{I}^{z}_{j,\bar{\sigma}} \right] 
   + \left[ J^z_{ij}- \frac{4}{9}V_{ij}(1-\xi) \right] \hat{S}^z_i\hat{S}^z_j \Bigg\},
\end{split}
\end{align}
where we have introduced the parameter~$\xi \in \mathbb{R}$ and for~$\xi=0$ we retrieve Eq.~\eqref{eq:tJ-XXZ}.
Here, we neglect the chemical potential terms; hence we assume that the driving frequency is low compared to the internal molecular energy scales~$\omega$ to suppress driving induced excitations, i.e.\,$1/T_F \ll \omega$.
On the other hand, we require the driving frequency to be much faster than flip-flop interactions,~$1/T_F \gg \chi$.
Since, $\omega \sim 1~\mathrm{GHz}$ and $\chi \sim 1~\mathrm{kHz}$ the limits can be achieved without concern of heating at this stage, or higher-order processes.

The above Hamiltonian can be derived as the effective Floquet model engineered from the underlying molecular Hamiltonian as we now demonstrate.
For two interacting levels, the molecular Hamiltonian (see main text) is given by
\begin{align} 
\begin{split} \label{eq:SuppHmol}
    \H_\mathrm{mol} = \sum_{i<j}\chi_{ij} \left(  \hat{I}^{x}_{i,\downarrow} \hat{I}^{x}_{j,\downarrow} + \hat{I}^{y}_{i,\downarrow} \hat{I}^{y}_{j,\downarrow}  \right),
\end{split}
\end{align} 
where flip-flop terms~$\chi_{ij}$ correspond to resonant dipole--dipole couplings~\cite{Wall2015}.

To obtain the effective Hamiltonian, we define global rotations on the~$\hat{\mathbf{S}}$-Bloch sphere by unitary operators~$\hat{U}^{\alpha}(\varphi)$, where~$\alpha=x,y,z$ is the rotation axis and~$\varphi$ the angle of rotation.
To first order in~$\tau/T_F$ and assuming instantaneous rotations, the effective Hamiltonian is given by
\begin{subequations}
\begin{align}
    \H_\mathrm{eff} &= \sum_{n} \dfrac{\tau_n}{T_F} \H_n \label{eq:effModel_sum} \\
    \H_n &= \hat{U}^\dagger_n \H_{\mathrm{mol}} \hat{U}_n,
\end{align}
\end{subequations}
where~$\tau_n$ is the evolution time of the $n$-th Floquet step, see Fig.~\ref{fig1}d, and $T_F = \sum_{n} \tau_n$ is the total time of one Floquet cycle. 
Moreover, we have defined~$\hat{U}_n$ to be the product of all rotations preceding the~$n$-th Floquet step.

\subsection{Floquet sequence without hole-hole density interaction~$V$} 

\begin{table}
\begin{center}
\renewcommand{\arraystretch}{1.5}
\newcolumntype{Y}{>{\centering\arraybackslash}X}
\begin{tabularx}{\linewidth}{|c|| Y | Y | Y | Y | Y | Y | Y | Y | Y |} 
 \hline
 time\,$\cdot\,T$ & $\tau_1$ & $\tau_2$ & $\tau_1$ & $\tau_3$ & $\tau_4$ & $\tau_3$ & $\tau_4$ & $\tau_3$ & $\tau_1$\\ [0.5ex] 
 \hline\hline 
 $\hat{S}^x_i \hat{S}^x_j$ & 0 & 0 & 0 & $\chi$ & 0 & $\chi$ & $\chi$ & $\chi$ & 0 \\ [0.5ex] 
 \hline
 $\hat{S}^y_i \hat{S}^y_j$ & 0 & 0 & 0 & $\chi$ & $\chi$ & $\chi$ & 0 & $\chi$ & 0 \\ [0.5ex] 
 \hline
 $\hat{S}^z_i \hat{S}^z_j$ & 0 & 0 & 0 & 0 & $\chi$ & 0 & $\chi$ & 0 & 0 \\ [0.5ex] 
 \hline
  $\hat{I}^x_{\downarrow,i} \hat{I}^x_{\downarrow,j}$ & $\chi$ & 0 & $\chi$ & 0 & 0 & 0 & 0 & 0 & $\chi$ \\ [0.5ex] 
 \hline
 $\hat{I}^y_{\downarrow,i} \hat{I}^y_{\downarrow,j}$ & $\chi$ & 0 & $\chi$ & 0 & 0 & 0 & 0 & 0 & $\chi$ \\ [0.5ex] 
 \hline
  $\hat{I}^z_{\downarrow,i} \hat{I}^z_{\downarrow,j}$ & 0 & 0 & 0 & 0 & 0 & 0 & 0 & 0 & 0 \\ [0.5ex] 
 \hline
    $\hat{I}^x_{\uparrow,i} \hat{I}^x_{\uparrow,j}$ & 0 & $\chi$ & 0 & 0 & 0 & 0 & 0 & 0 & 0\\ [0.5ex] 
 \hline
     $\hat{I}^y_{\uparrow,i} \hat{I}^y_{\uparrow,j}$ & 0 & $\chi$ & 0 & 0 & 0 & 0 & 0 & 0 & 0 \\ [0.5ex] 
 \hline
   $\hat{I}^z_{\uparrow,i} \hat{I}^z_{\uparrow,j}$ & 0 & 0 & 0 & 0 & 0 & 0 & 0 & 0 & 0 \\ [0.5ex] 
 \hline
\end{tabularx}
\end{center}
\caption{\textbf{Effective Floquet couplings with $V=0$.} Global rotations of the molecular Hamiltonian~\eqref{eq:molHam} generate the terms listed in the first column. The rotations and evolution steps~$\tau_n$ correspond to the sequence shown in Fig.~\ref{fig1}d and Fig.~\ref{figSupp_Floquet_2level}c. The coupling strength arise in the summands of Eq.~\eqref{eq:effModel_sum} and have to be multiplied by~$\tau_n/T$. Here we have suppressed site indices.}
\label{tab:effCouplings_noV}
\end{table}

We propose a sequence of microwave rotations that average to an effective Floquet time evolution under a \mbox{$t$\,--\,$J$~model} without hole-hole interaction~$V=0$, see Fig.~\ref{figSupp_Floquet_2level}b.
The form and coupling strengths of Hamiltonian~\eqref{eq:effModel_sum} are summarized in Tab.~\ref{tab:effCouplings_noV}.
Enforcing the constraints that hopping of~$\downarrow$- and $\uparrow$-particles should have equal amplitudes as well as equal magnetic XX~and YY~interactions constrains the time steps in the Floquet evolution.
Therefore, we obtain the following set of equations:

\begin{subequations} \label{eq:taus_withoutV}
\begin{align} 
    \frac{\tau}{T} &= \left( 6\epsilon^\perp + 24\epsilon^t + 3 \right)^{-1} \\
    \tau_1 &= 4\epsilon^t\cdot\tau \\
    \tau_2 &= 12\epsilon^t\cdot\tau \\
    \tau_3 &= \tau \\
    \tau_4 &= 3\tau
\end{align}
\end{subequations}  
with the effective coupling strength
\begin{align}
    \frac{J^z}{\chi} = \left(\epsilon^\perp + 4\epsilon^t+\frac{1}{2}\right)^{-1}.
\end{align}

To justify the first-order Floquet expansion, we perform exact diagonalization studies including long-ranged dipolar interactions and finite pulse width, i.e. finite Rabi frequency~$\Omega$; for simplicity we assume the same Rabi frequency for all three transitions shown in Fig.~\ref{figSupp_Floquet_2level}a.
We compare the time evolution under the Floquet time evolution with the theoretically predicted target model for~$t=J^\perp=J^z=1$ and $V=0$.

\begin{figure}[t]
\centering
\includegraphics[width=0.95\linewidth]{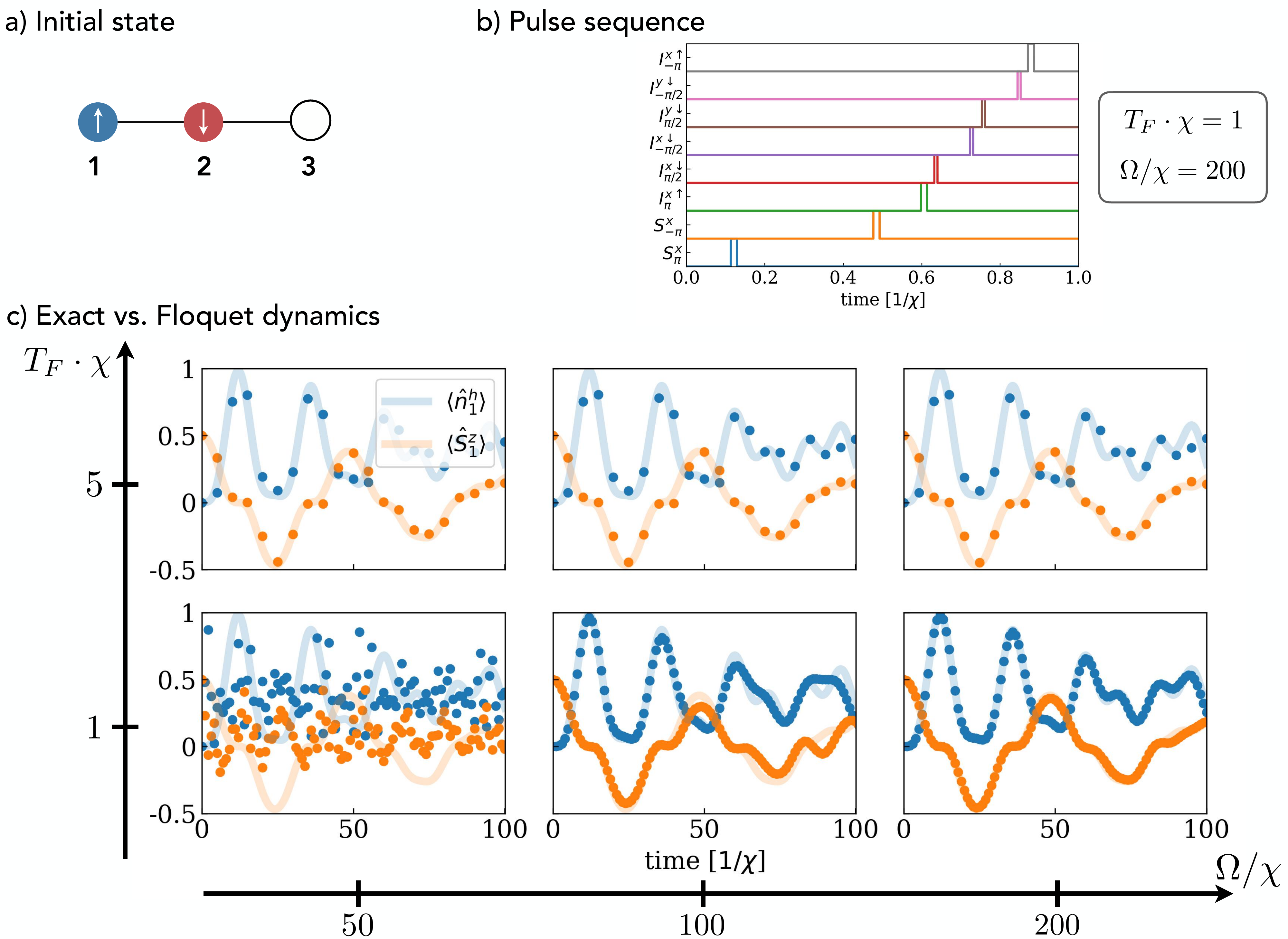}
\caption{\textbf{Floquet time evolution.} \textbf{a)} We perform exact time evolution of a system of three molecules by quenching the initial state. \textbf{b)} The system evolves under the molecular Hamiltonian and under a sequence of finite width Rabi pulses. The sequence gives rise to a \mbox{$t$\,--\,$J$~model} with~$t/J=1$ ($t=2/11\chi$) in first-order Floquet theory. \textbf{c)} We compare the stroboscopic (dots) to exact target \mbox{$t$\,--\,$J$} (solid) time evolution the hole occupation (blue) and magnetization (orange) on site~$1$. We vary the Floquet cycle time~$T_F$ as well as the Rabi frequency~$\Omega$. For~$\Omega/\chi=50$ and~$T_F\chi=1$, the Rabi pulses overlap such that the Floquet prediction is invalid. For all other parameter regimes, we find very robust prediction of the dynamics. }
\label{figSupp_Floquet}
\end{figure}

In our numerical calculations, we initialize a system of three molecules in a product state~$\ket{\psi(t=0)} = \ket{\uparrow}_1\otimes\ket{\downarrow}_2\otimes\ket{h}_3$, see Fig.~\ref{figSupp_Floquet}a.
We continuously time evolve the system under Hamiltonian~\eqref{eq:SuppHmol} (nearest-neighbour interaction strength~$\chi$) and apply periodic rotations between the three levels according to the sequence shown in Fig.~\ref{figSupp_Floquet}b with various realistic Rabi frequencies~$\Omega/\chi= 50,100,200$. Moreover, we vary the time~$T_F$ of a single Floquet cycle.
Finally, we stroboscopically measure at times~$t_M = M T_F$ with~$M \in \mathbb{N}$ and compare observables to the time evolution of an exact \mbox{$t$\,--\,$J$~model} shown in Fig.~\ref{figSupp_Floquet}c.
We find excellent agreement over the entire range of parameters demonstrating the robustness of the Floquet scheme.
Note that for very short times~$T_F$ and slow Rabi frequencies~$\Omega$, the Floquet prediction breaks down due to overlapping microwave pulses. As soon as the individual pulses are separated, the time evolution is well described by the first-order Floquet Hamiltonian with increasing fidelity for shorter~$T_F$ and faster~$\Omega$.

\subsection{Floquet sequence with hole-hole density interaction~$V$} 

As discussed in the main text, an interesting parameter to tune is the hole-hole interaction~$\propto V$, which can be achieve using the proposed Floquet sequence shown in Fig.~\ref{figSupp_Floquet_2level}c. We target a $t$\,-$J$\,-\,$V$~model as written in Eq.~\eqref{eq:tJ-withXi} with~$\xi=0$.
From Tab.~\ref{tab:effCouplings_withV}, we can derive the required Floquet times for models with~$0 < \epsilon^V < 9/4$ and~$\epsilon^\perp > 1/2$:

\begin{subequations} \label{eq:taus_withV}
\begin{align}
    \frac{\tau}{T} &= \left[ \left(\epsilon^\perp + 2\epsilon^t + 2\right)\frac{\tau_3}{\tau} +2(2\epsilon^t + \epsilon^\perp) - 1  \right]^{-1} \\
    \tau_1 &= \left[\frac{\tau_3}{5\tau}(\epsilon^t+1) + \frac{2}{5}\epsilon^t \right]\cdot\tau \\
    \tau_2 &= \frac{5}{3}\cdot\tau_1 \\
    \tau_3 &= \frac{8}{9/\epsilon^V-4}\cdot\tau \\
    \tau_4 &= \frac{1}{3}\left[ \epsilon^\perp\left( 2 + \frac{\tau_3}{\tau}\right) - 1 \right]\cdot\tau \\
    \tau_5 &= \tau
\end{align}
\end{subequations}  
with the effective coupling strength
\begin{align}
    \frac{J^z}{\chi} = \frac{2\tau+\tau_3}{T}.
\end{align}

\begin{table}
\begin{center}
\renewcommand{\arraystretch}{1.5}
\newcolumntype{Y}{>{\centering\arraybackslash}X}
\begin{tabularx}{\linewidth}{|c|| Y | Y | Y | Y | Y | Y | Y | Y | Y | Y | Y | Y | Y | Y | Y | Y | Y | Y | Y |} 
 \hline
 time\,$\cdot\,T$ & $\tau_1$ & $\tau_2$ & $\tau_3$ & $\tau_2$ & $\tau_3$ & $\tau_2$ & $\tau_1$ & $\tau_3$ & $\tau_1$ & $\tau_3$ & $\tau_1$ & $\tau_4$ & $\tau_5$ & $\tau_4$ & $\tau_5$ & $\tau_4$ & $\tau_1$\\ [0.5ex] 
 \hline\hline 
 $\hat{S}^x_i \hat{S}^x_j$ & 0 & 0 & 0 & 0 & 0 & 0 & 0 & 0 & 0 & 0 & 0 & $\chi$ & 0 & $\chi$ & $\chi$ & $\chi$ & 0 \\ [0.5ex] 
 \hline
 $\hat{S}^y_i \hat{S}^y_j$ & 0 & 0 & 0 & 0 & 0 & 0 & 0 & 0 & 0 & 0 & 0 & $\chi$ & $\chi$ & $\chi$ & 0 & $\chi$ & 0  \\ [0.5ex] 
 \hline
 $\hat{S}^z_i \hat{S}^z_j$ & 0 & 0 & 0 & 0 & 0 & 0 & 0 & 0 & 0 & 0 & 0 & 0 & $\chi$ & 0 & $\chi$ & 0 & 0 \\ [0.5ex] 
 \hline
  $\hat{I}^x_{\downarrow,i} \hat{I}^x_{\downarrow,j}$ & $\chi$ & 0 & 0 & 0 & 0 & 0 & $\chi$ & $\chi$ & $\chi$ & 0 & $\chi$ & 0 & 0 & 0 & 0 & 0 & $\chi$  \\ [0.5ex] 
 \hline
 $\hat{I}^y_{\downarrow,i} \hat{I}^y_{\downarrow,j}$ & $\chi$ & 0 & 0 & 0 & 0 & 0 & $\chi$ & 0 & $\chi$ & $\chi$ & $\chi$ & 0 & 0 & 0 & 0 & 0 & $\chi$  \\ [0.5ex] 
 \hline
  $\hat{I}^z_{\downarrow,i} \hat{I}^z_{\downarrow,j}$ & 0 & 0 & 0 & 0 & 0 & 0 & 0 & $\chi$ & 0 & $\chi$ & 0 & 0 & 0 & 0 & 0 & 0 & 0  \\ [0.5ex] 
 \hline
    $\hat{I}^x_{\uparrow,i} \hat{I}^x_{\uparrow,j}$ & 0 & $\chi$ & 0 & $\chi$ & $\chi$ & $\chi$ & 0 & 0 & 0 & 0 & 0 & 0 & 0 & 0 & 0 & 0 & 0 \\ [0.5ex] 
 \hline
     $\hat{I}^y_{\uparrow,i} \hat{I}^y_{\uparrow,j}$ & 0 & $\chi$ & $\chi$ & $\chi$ & 0 & $\chi$ & 0 & 0 & 0 & 0 & 0 & 0 & 0 & 0 & 0 & 0 & 0 \\ [0.5ex] 
 \hline
   $\hat{I}^z_{\uparrow,i} \hat{I}^z_{\uparrow,j}$ & 0 & 0 & $\chi$ & 0 & $\chi$ & 0 & 0 & 0 & 0 & 0 & 0 & 0 & 0 & 0 & 0 & 0 & 0 \\ [0.5ex] 
 \hline
\end{tabularx}
\end{center}
\caption{\textbf{Effective Floquet couplings with $V$.} Global rotations of the molecular Hamiltonian~\eqref{eq:molHam} generate the terms listed in the first column. The rotations and evolution steps~$\tau_n$ correspond to the sequence shown in Fig.~\ref{figSupp_Floquet_2level}c. The coupling strength arise in the summands of Eq.~\eqref{eq:effModel_sum} and have to be multiplied by~$\tau_n/T$. Here we have suppressed site indices.}
\label{tab:effCouplings_withV}
\end{table}

\section{Ground-state calculations with DMRG}

We calculate the 0-, 1- and 2-hole ground states of the Hamiltonian \eqref{eq:tJ-general} on an $L_x \times L_y$ square lattice with cylindrical boundary conditions (CBCs) using the density matrix renormalization group (DMRG) algorithm \cite{White1992,Schollwoeck2011,Hubig2017,Syten}. Specifically, we implement the following $t$\,--\,$J$ Hamiltonian

\begin{equation}
\label{eq:tJ-DMRG}
    \H_{t-J} = -t \sum_{\ij\sigma} \hat{\mathcal{P}}\left( \a^\dagger_{i,\sigma}\a^{\phantom{\dagger}}_{j,\sigma} + \mathrm{h.c.}  \right) \hat{\mathcal{P}}
    + J\sum_{\braket{i,j}} \left(\hat{\bm{S}}_i\cdot \hat{\bm{S}}_j  - \frac{1}{4} \n_i \n_j\right),
\end{equation}

\noindent where $\a^{\dagger}_{i,\sigma}, \a^{\phantom\dagger}_{i,\sigma}$ are the bosonic (fermionic) creation and annihilation operators satisfying canonical (anti-)commutation relations, and $\langle i,j \rangle$ denotes a sum over all nearest-neighbour (NN) pairs of sites; $t=t_{ij}$ is the NN tunnelling amplitude, $J=J^\alpha$ $(\alpha=x,y,z)$ is the isotropic Heisenberg exchange coupling, and $\hat{\mathcal{P}}$ is the Gutzwiller operator which projects out all states with double occupancies. Simulations are performed for our chosen parameters $t/J=2$ and $V/J =-1/4$, which may be experimentally realized in bosonic systems via the Floquet engineering scheme outlined above. 

We explicitly implement a global U$(1)_N \otimes $U$(1)_{S_z}$ symmetry -- corresponding to conserved total particle number and magnetization -- to improve the efficiency of the DMRG algorithm. Our calculations utilize bond dimensions up to $m\sim 6000$, and for the two hole case we obtain final truncation errors $\mathcal{T}(\varepsilon)\sim 1\times 10^{-6}$. To ensure that the obtained ground states are well converged, we monitor local one- and two-site observables, including the local hole density $\braket{\hat{n}^h_i}$ and spin expectation values $\braket{\hat{S}^\alpha_i}$, as well as the reduced hole-hole $\braket{\hat{n}^h_i\hat{n}^h_j}/\braket{\hat{n}^h_i}\braket{\hat{n}^h_j}$ and spin-spin correlation functions $\braket{\hat{S}^\alpha_i\hat{S}^\beta_j}$. In particular, we find that $|\langle \hat{S}^z_{i}\rangle|\sim\mathcal{O}(10^{-7})$ and  $|4\langle \hat{S}^z_{i}\hat{S}^z_{j} \rangle - \langle \hat{S}^+_{i}\hat{S}^-_{j} + \mathrm{h.c.}\rangle|\sim \mathcal{O}(10^{-6})$, which should both vanish identically due to the global SU(2)-spin symmetry of our model. 

\begin{figure}[t]
\centering
\includegraphics[width=0.6\linewidth]{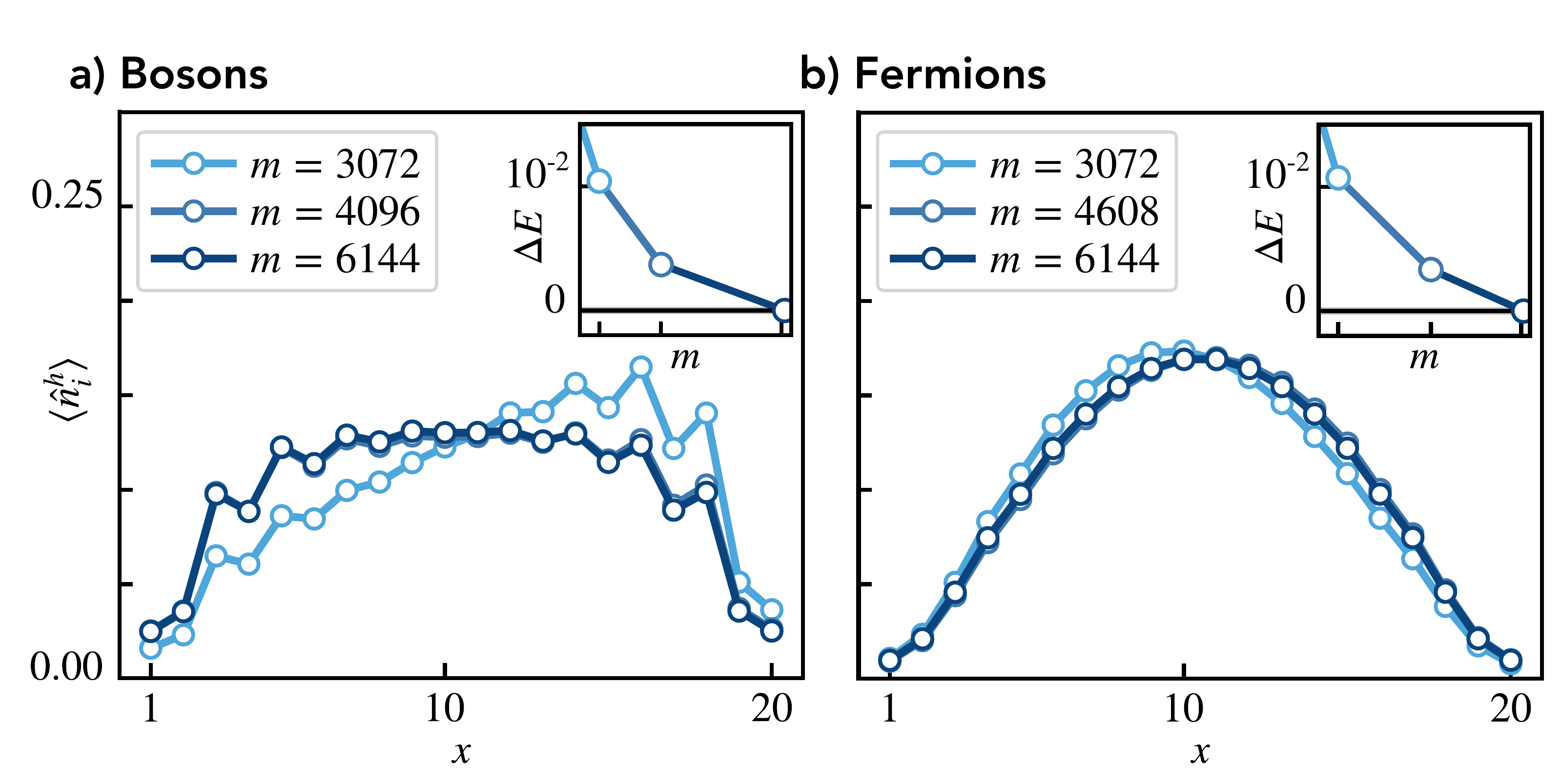}
\caption{\textbf{DMRG convergence.} Convergence of the local hole-density $\braket{\hat{n}^h_i}$ with the bond dimension $m$ of the matrix product state. Data is shown for both the bosonic \textbf{a)} and fermionic \textbf{b)} ground states, each for two holes on a $20 \times 6$ cylinder. The hole densities are summed around the cylinder and displayed as a function of the coordinate $x$ running along the long direction. The plots for $m \gtrsim 4000$ overlap almost perfectly, independent of statistics. The insets show the convergence of the energy $E$ with bond dimension $m$. Here, $\Delta E$ is the difference between the energy at bond dimension $m$ relative to to the final energy at $m_{\mathrm{max}}$. The $3$ points show the same values of the bond dimension as the main panels. }
\label{figSupp_convergence}
\end{figure}

In Fig. \ref{figSupp_convergence}, we plot the expectation value of the local hole density along the long direction of the $20 \times 6$ cylinder for different values of the bond dimension. The hole density serves as a sensitive probe for convergence, as slight real-space shifts of the single hole pair delocalized over an extensive lattice come with a very small energy-cost. Independent of statistics, $\braket{\hat{n}_i^h}$ is still noticeably asymmetric for $m \sim 3000$, but symmetric and almost identical for values $m \gtrsim 4000$. With this most sensitive probe, we conclude our analysis of DMRG convergence.

\end{document}